\documentclass[runningheads]{llncs}
\usepackage[T1]{fontenc}
\usepackage{graphicx}
\usepackage{color}
\usepackage{url}
\usepackage{bm}
\usepackage{amsmath}
\usepackage{amssymb}
\usepackage[caption=false]{subfig}

\begin{document}
\title{Efficient Neural Generation of 4K Masks for Homogeneous Diffusion 
Inpainting 
\thanks{This work has received funding from the European Research
Council (ERC) under the European Union's Horizon 2020 research and innovation 
programme (grant agreement no. 741215, ERC Advanced Grant INCOVID).}
}
\titlerunning{Neural Generation of 4K Masks for Homogeneous Diffusion 
Inpainting}
\author{Karl Schrader\inst{1}%
\and
Pascal Peter\inst{1}%
\and
Niklas K\"amper\inst{1}%
\and
Joachim Weickert\inst{1}%
}
\authorrunning{K. Schrader, P. Peter, N. K\"amper, J. Weickert}
\institute{Mathematical Image Analysis Group,
Faculty of Mathematics and Computer Science, Campus
E1.7, Saarland University, 66041 Saarbr\"ucken, Germany.
\email{\{schrader,peter,kaemper,weickert\}@mia.uni-saarland.de}%
}
\maketitle              %
\begin{abstract}
With well-selected data, homogeneous diffusion inpainting can reconstruct 
images from sparse data with high quality. While 4K colour images of size 
$3840 \times 2160$ can already be inpainted in real time, optimising the known 
data for applications like image compression remains challenging: Widely used 
stochastic strategies can take days for a single 4K image.
Recently, a first neural approach for this so-called mask optimisation problem 
offered high speed and good quality for small images. It trains a mask 
generation network with the help of a neural inpainting surrogate.
However, these mask networks can only output masks for the resolution and 
mask density they were trained for.
We solve these problems and enable mask optimisation for high-resolution 
images through a neuroexplicit coarse-to-fine strategy. Additionally, we 
improve the training and interpretability of mask networks by including a 
numerical inpainting solver directly into the network. This allows to generate 
masks for 4K images in around $0.6$ seconds while exceeding the quality of 
stochastic methods on practically relevant densities. Compared to popular
existing approaches, this is an acceleration of up to four orders of
magnitude.

\end{abstract}
\section{Introduction}

Inpainting-based image compression \cite{GWWB08} is surprisingly 
simple: During encoding, only a carefully 
optimised subset of all pixel locations and their values is stored.
In the decoding phase, the missing information is reconstructed in a
lossy way by some inpainting method. Optimising the inpainting data,
the so-called mask, is essential for obtaining a good reconstruction: 
Even with a simple method such as homogeneous diffusion inpainting 
\cite{Ca88}, results can be surprisingly good.

\smallskip
Simple model-driven approaches for this optimisation problem require a 
compromise between reconstruction quality and speed~\cite{BBBW08,MHWT12}. 
Fast and qualitatively convincing results rely on sophisticated algorithms 
and implementations~\cite{CW21}.
A first attempt to combine both advantages within a deep learning
setting has been proposed recently by Alt et al.~\cite{APW22}. It 
is simple, fast, and approaches the quality of widely used 
model-driven optimisation strategies. It trains a mask generator 
network with the help of a neural inpainting surrogate. 
Extensions~\cite{PSAW22} allow to optimise not only the 
positions, but also the pixel values of the known data.

\smallskip
While offering a very good combination of quality and speed for 
greyscale images of sizes up to $256\times256$ and colour images up to 
$128\times128$, it is inflexible: 
The mask network can only generate masks for the density and resolution it was 
trained for. Furthermore, a na\"ive extension to high-resolution images 
requires a prohibitive amounts of compute power during training: 
At a resolution of $3840\times2160$, 4K colour images have around $25$
million values, two to three orders of magnitude more than previously 
considered.

\medskip\noindent
\textbf{Our Contribution.}
With the first coarse-to-fine approach for neural mask generation, 
we address the weak points of the neural approach while preserving 
its high speed and increasing the quality of the generated masks.
We partition the image into patches, and generate masks for each. 
Mask pixel budgets are assigned to each patch using an optimality result 
of Belhachmi et al.~\cite{BBBW08}. This constitutes the first transfer of 
their findings to the discrete setting which does not involve dithering,
a step that usually leads to substantial quality losses. 
Our approach matches the quality of widely used stochastic mask optimisation 
strategies on 4K images of size $3840 \times 2160$, while being up to four 
orders of magnitude faster.
With a new training process, we solve the challenging problem of integrating 
model-based inpainting directly into the network. This removes the previous 
need for surrogate solvers and yields an overall more transparent and reliable 
architecture. Additionally, it greatly reduces the number of trainable weights 
and speeds up the training.

\medskip\noindent
\textbf{Related Work.}
There are many approaches for mask optimisation. The taxonomy below is an 
extension of the one of Peter et al.~\cite{PSAW22}.
\begin{enumerate}
\item \textit{Analytic Approaches.} 
For the continuous setting, Belhachmi et al.~\cite{BBBW08} were able to show 
that optimal masks can be derived from the Laplacian magnitude (modulus) of the 
image. To discretise this result, the Laplacian magnitude has been dithered so 
far. We apply this result in a novel way to estimate optimal patch densities in 
Section \ref{sec:4k_mask_gen}.
Since analytic approaches do not require any inpaintings, they are very
fast. However, dithering comes with significant compromises and typically
results in low quality~\cite{MHWT12}.\vspace*{1mm}

\item \textit{Nonsmooth Optimisation Strategies.} 
Finding a good inpainting mask that minimises the mean squared error of the 
reconstruction can be phrased as a nonsmooth bilevel optimisation problem. It 
can be solved with primal-dual methods~\cite{BLPP16,HSW13,OCBP14}.
While being able to achieve high quality, they do not allow the user to specify 
a target density as for to our approach. In addition, they produce nonbinary 
masks and require binarisation as a postprocessing step which leads to a loss 
of quality. This can be alleviated by applying tonal optimisation 
\cite{MHWT12}, which optimises not only the position, but also the value of the 
mask pixels. While this can introduce small errors to the known data, it is 
often beneficial for the overall quality of the reconstruction.

\item \textit{Sparsification Methods.} 
This class of methods starts with a full mask and successively remove mask 
pixels with a low impact on reconstruction quality. One widely used 
implementation is \textit{probabilistic sparsification} by 
Mainberger et al.~\cite{MHWT12}.
It allows a target density to be specified and produces binary masks directly. 
However, it requires an inpainting to be computed in every step which results 
in long runtimes, especially on large images.

\item \textit{Densification Methods.} 
They start with an empty mask and successively add those pixels to it which 
increase reconstruction quality the most~\cite{DAW21,GWWB08}. 
Combined with ideas like finite elements~\cite{CW21}, they constitute a 
fairly new but promising class of methods.

\item \textit{Relocation Methods.} 
Greedy optimisation strategies can get stuck in local minima. 
To escape from them, \textit{nonlocal pixel exchange}~\cite{MHWT12} tests 
if moving some randomly selected mask pixels into the unknown regions 
leads to a better reconstruction. Successful moves are kept, while 
unsuccessful ones are reverted. Given sufficient time, this method can 
produce very good masks. It is usually applied as a postprocessing tool.

\item \textit{Neural Methods.} 
Works such as~\cite{DCPC19,Pe22} learn the inpainting operator along with a 
mask generation network. As this results in an opaque model, we focus on 
well-understood homogeneous diffusion inpainting. 
For it,~\cite{APW22,PSAW22} have shown that a mask 
generation network can be successfully trained with the help of an inpainting 
surrogate, but focus on small images of sizes up to $256\times256$. 
Opposed to our model, new densities and resolutions require a new model to be 
trained. Their framework also allows for the training of a 
tonal optimisation network with the same strategy. While this is also the case 
for our approach, we focus on the spatial case due to space limitations. 
Inference with neural mask generation is very fast since it requires no 
inpaintings.
\end{enumerate}
These approaches can be summarised as follows:
Category 1 is very fast as it requires no inpaintings, but offers low quality. 
Higher quality is be achieved by Categories 2 -- 4 at the expense of speed,
as they all need to calculate many inpaintings. Given sufficient iterations, 
Category~5 can improve any mask further. Previous neural methods 
from Category 6 offer good quality and speed, but are not as flexible 
as model-driven approaches. We will exploit optimality results from Category 1 
without sacrificing quality through dithering to alleviate the shortcomings of 
Category 6.

\medskip \noindent
\textbf{Organisation of the Paper.}
Section~\ref{sec:spatial_optimisation} reviews inpainting with homogeneous 
diffusion and model-driven spatial optimisation. 
Section \ref{sec:neural_mask_generation} presents our improvements
to the baseline neural mask generation architecture, followed by our new 
coarse-to-fine mask generation strategy in Section~\ref{sec:4k_mask_gen}. 
Section~\ref{sec:experiments} provides empirical evidence for our patch density 
estimation and compares our neural approach against model-driven ones. 
We end with conclusions and an outlook in Section~\ref{sec:conclusion}.

\section{Homogeneous Diffusion Inpainting} \label{sec:spatial_optimisation}
Let us consider a continuous greyscale image 
$f: \Omega \to \mathbb{R}$ defined on a rectangular image domain $\Omega$ 
which is only known on a mask $K \subset \Omega$. The goal of 
inpainting is to restore the inpainting region $\Omega \setminus K$ 
with a homogeneous diffusion process that propagates information; see 
e.g.~\cite{GL14}. For homogeneous diffusion inpainting~\cite{Ca88}, the 
reconstruction solves the partial differential equation 
\begin{equation}
(1-c) \Delta u - c(u-f) = 0 \label{eq:hd_inpainting}
\end{equation}
where $c:\Omega\to\{0, 1\}$ is a binary confidence function which indicates 
if a point $\bm x$ is part of the mask. Known values are specified by 
$c(\bm x) = 1$ and are preserved. Missing data are indicated by 
$c(\bm x) = 0$ and are inpainted with the steady state $\Delta u =0$ of the 
homogeneous diffusion equation, where $\Delta = \partial_{xx} + \partial_{yy}$ 
is the Laplacian. 
We assume reflecting boundary conditions at the image boundary 
$\partial \Omega$. Nonbinary masks are also well-defined within 
the mathematical framework, but are hard to compress and only used as 
an intermediate step in this work.
For colour images, each channel is considered independently, since 
homogeneous diffusion is a linear operator.

\smallskip
While more sophisticated inpainting models exist~\cite{GWWB08,WW06}, 
homogeneous diffusion is attractive for inpainting-based compression: 
Its simplicity allows for an analytic continuous theory for inpainting 
data selection~\cite{BBBW08} and enables fast implementations~\cite{KW22}. 
Furthermore, its complete lack of parameters make it easy to use, and it can 
yield high-quality reconstruction for well optimised inpainting data. 
These properties make it a suitable candidate for mask optimisation.

\smallskip
In the discrete setting, we consider digital colour images 
$\bm f \in \mathbb{R}^{3n_xn_y}$ with resolution $n_x \times n_y$. 
The inpainting equation \eqref{eq:hd_inpainting} is discretised with finite 
differences, leading to a linear system of equations. The reconstruction 
$\bm u$ can then be computed by a suitable numerical solver 
such as the conjugate gradient method; see e.g.~\cite{We17}. Mask optimisation 
strategies 
are designed to generate discrete masks~$\bm c \in \{0, 1\}^{n_xn_y}$ such that 
the corresponding reconstruction $\bm u$ approximates the original image well. 
They are constrained by the mask density $\|\bm c\|_1/(n_xn_y)$ where $\|.\|_1$ 
is the $1$-norm, which should match the desired target density $d$.

\subsection{Model-Based Mask Optimisation}
A mask optimisation strategy that does not require any inpaintings has been 
proposed by Belhachmi et al.~\cite{BBBW08}. They show that for optimal masks,  
the local density should increase with the Laplacian magnitude. In 
Section~\ref{sec:4k_mask_gen} we use this result to predict a suitable 
density for image regions. 
However, the application of these results to discrete images relies on 
dithering. The commonly used Floyd-Steinberg dithering~\cite{FS76} is 
fast and simple, but suffers from a directional bias and generates masks of
relatively low quality.

\smallskip
A method which produces better masks but remains simple is 
\textit{probabilistic sparsification}~\cite{MHWT12}. 
It starts with a full mask and gradually removes pixels until the target
density $d$ is reached. In every iteration, a fraction $p$ of mask pixels is 
removed. Then, an inpainting is computed, and the fraction $q$ which resulted 
in the largest loss of quality is added back. 

\smallskip
We also consider \textit{nonlocal pixel exchange}~\cite{MHWT12} as a 
post-processing method. In each step, it moves a fraction $p$ of mask pixels 
into the unknown image area. It then inpaints with the new mask to check for 
improvements: Exchanges which increase inpainting quality are kept, 
while unsuccessful ones are reverted. 
This is repeated for $k$ cycles with $\|\bm c\|_1$ iterations each.
Nonlocal pixel exchange can help greedy approaches like probabilistic 
sparsification escape from poor local minima at the cost of significant 
runtime.

\smallskip
Both probablisitic sparsification and nonlocal pixel exchange require an 
inpainting for each iteration, with the other operations being 
quick in comparison. As such, their runtime is primarily determined by 
the number and the speed of the inpainting operations. 
A na\"ive CPU-based inpainter using a conjugate gradient solver 
takes around $3.5$ seconds per 4K colour image on a contemporary PC. 
At this speed, 5 cycles of nonlocal pixel exchange for a mask with $5\%$ known 
data would take around $12$ weeks.
To enable meaningful comparisons, we have developed implementations 
of probablistic sparsification and nonlocal pixel exchange that use
the very fast GPU-based inpainting of K\"amper et al.~\cite{KW22}.
With it, an iteration of either method requires only $\approx 6$~milliseconds, 
a speedup of two orders of magnitude. To the best of our knowledge, 
our implementations are the fastest available.

\section{Neural Mask Generation}\label{sec:neural_mask_generation}
\begin{figure}[t]
\centering
\includegraphics[width=\linewidth]{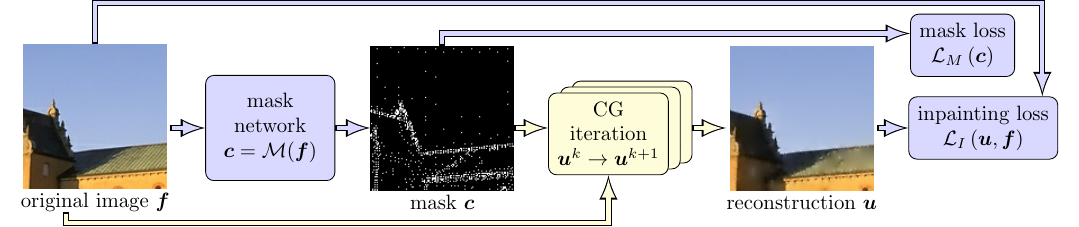}
\caption{\textbf{Our Mask Optimisation Architecture.} 
The mask network and its losses are coloured in blue, 
our newly introduced sequence of CG iterations is given in yellow. 
}
\label{fig:network_architecture}
\end{figure}

We follow the basic network architecture from Peter et al.~\cite{PSAW22} with a 
mask generator network and an inpainting approximator. 
The latter is used to train the mask network by evaluating 
the reconstruction quality and is discarded during inference. An overview of 
our architecture can be found in Figure \ref{fig:network_architecture}. 
While we retain the mask network as is, we completely replace the original 
inpainting approximator.

\medskip \noindent
\textbf{Mask Network.}
The mask network receives the original image $\bm f$ and generates a 
mask $\bm c = \mathcal{M}(\bm f)$. Its output is restricted to $[0, 1]$ by 
applying a sigmoid activation function. Additionally, the network outputs are 
rescaled if they exceed the desired density $d$.
The goal of the network is to produce masks such that the corresponding 
inpainted image $\bm u$ is close to the original $\bm f$. 
To this end, we minimise their mean squared error (MSE)
$
\mathcal{L}_I(\bm u, \bm f) = \frac{1}{n_x n_y}\|\bm u - \bm f\|_2^2
$
where $\|.\|_2$ is the Euclidean norm.
In addition, we require a mechanism to encourage the generation of binary 
masks. To this end, we penalise the inverse variance of the mask using 
$
\mathcal{L}_M(\bm c) = \alpha (\sigma_{\bm c}^2 + \varepsilon)^{-1}
$
where $\varepsilon$ is a small numerical constant to avoid division by $0$, and 
$\alpha$ balances variance loss $\mathcal{L}_M$ and inpainting loss 
$\mathcal{L}_I$.

\medskip \noindent
\textbf{Inpainting Approximator.}
To train the mask network with an inpainting loss~$\mathcal{L}_I$, we need to 
approximate the inpainting process inside the network. 
Previous approaches \cite{APW22,PSAW22} facilitate this by including a 
surrogate inpainting network which receives the original $\bm f$ and 
mask $\bm c$, and is trained to find a reconstruction $\bm u$ which 
solves the discrete version of the inpainting equation \eqref{eq:hd_inpainting}:
\begin{equation}
(\bm I - \bm C) \bm A \bm u - \bm C (\bm u - \bm f) = \bm 0 
\label{eq:discrete_inp}
\end{equation}
Here, the matrix $\bm C = \text{diag}(\bm c)$ contains the mask entries on the 
diagonal, and $\bm A$ applies a finite difference discretisation of the 
Laplacian $\Delta$ with reflecting boundary conditions.
This approach significantly increases the total number of weights and adds 
complexity to the architecture. Furthermore, it decreases interpretability as 
the surrogate is, apart from its loss function, a black box.

\smallskip
We propose to replace it by a sufficient number of iterations of a successful
numerical solver. As homogeneous diffusion leads to a linear system of 
equations with a symmetric system matrix, we can use the 
conjugate gradient~(CG) method~\cite{We17}. It offers convergence guarantees 
while remaining simple and efficient. As each iteration is differentiable, 
we can backpropagate through the solver to train the mask network. 
By introducing this well-understood numerical solver, we have reduced the 
total number of weights by half compared to~\cite{PSAW22} while increasing 
interpretability. Section \ref{sec:inpainting_approx} confirms that this 
improves inpainting approximation quality greatly and the quality 
of generated masks slightly.

\medskip \noindent
\textbf{Mask Generation in Practice.}
As the generated masks are not guaranteed to be binary, we need to binarise 
them. To this end, Peter et al.~\cite{PSAW22} perform weighted coinflips at 
each mask pixel and then choose the best-performing mask out of $30$ attempts. 
We found that rounding leads to a comparable quality for our masks. It is less 
time intensive and involves no randomness.

\section{Coarse-To-Fine Approach for Mask Generation} \label{sec:4k_mask_gen}
\begin{figure}[t]
\begin{tabular}{cc}
(a) initial image & (b) quantised patch densities in \%\\
\includegraphics[width=6cm]{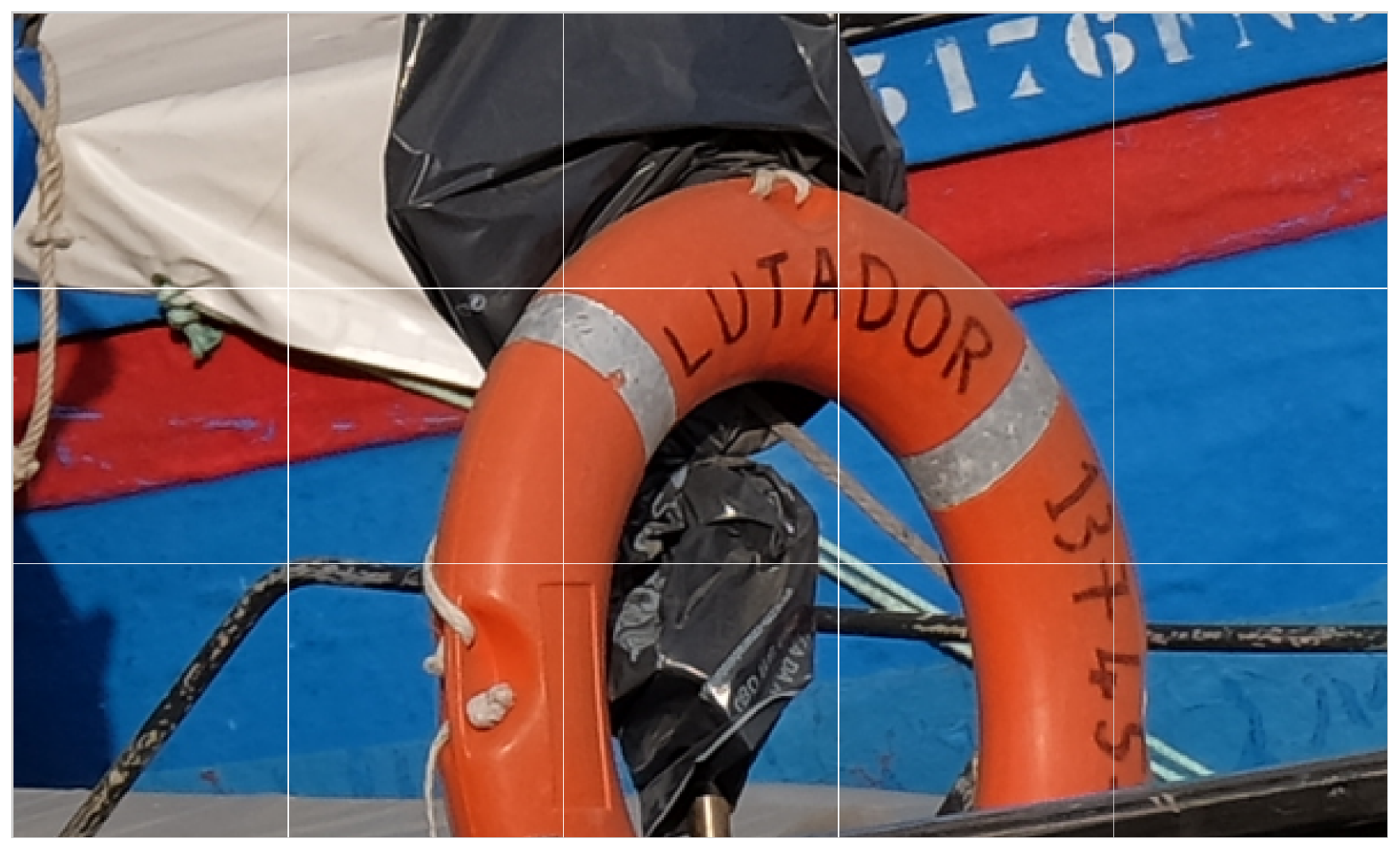} &
\includegraphics[width=6cm]{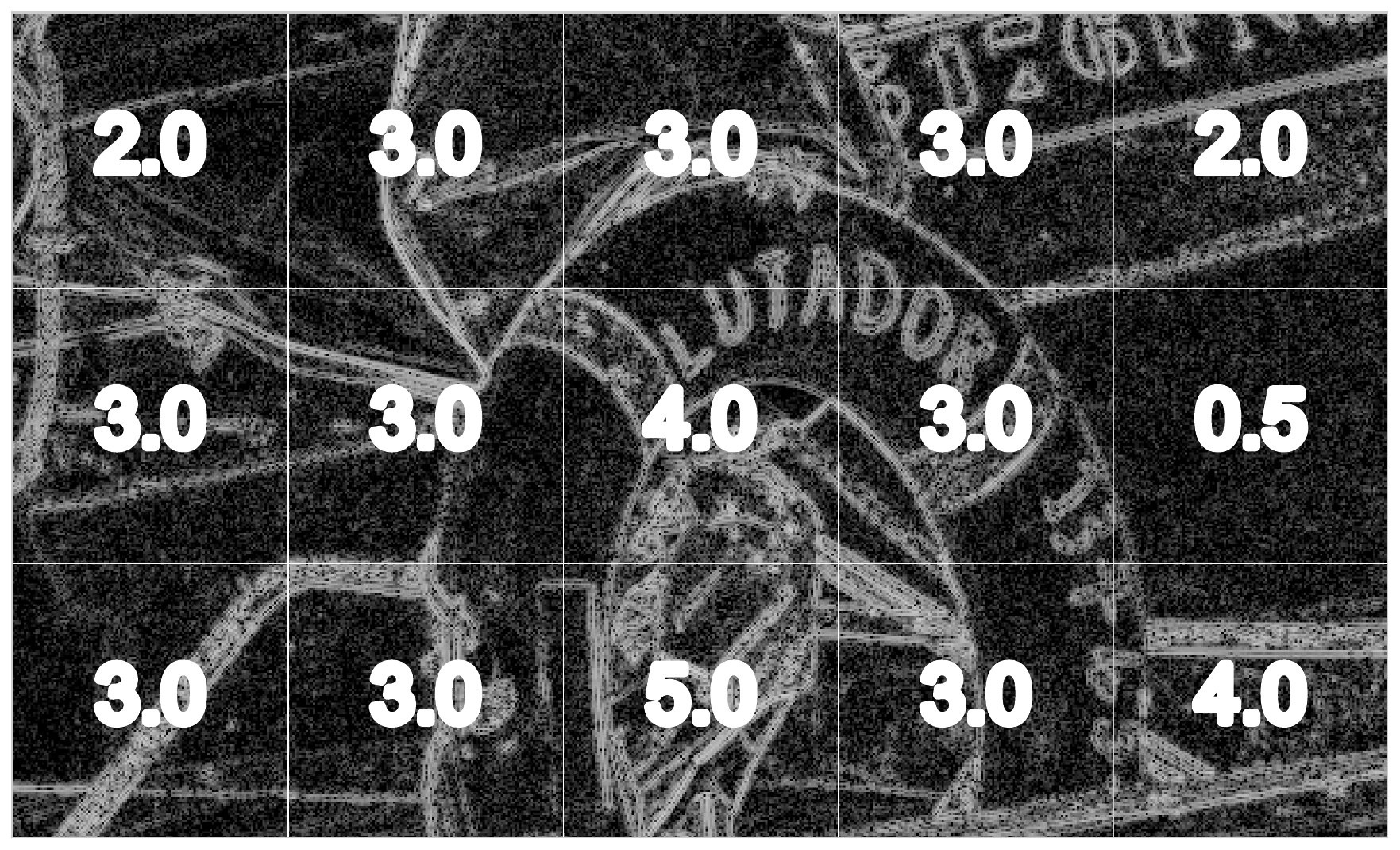} \\
\includegraphics[width=6cm]{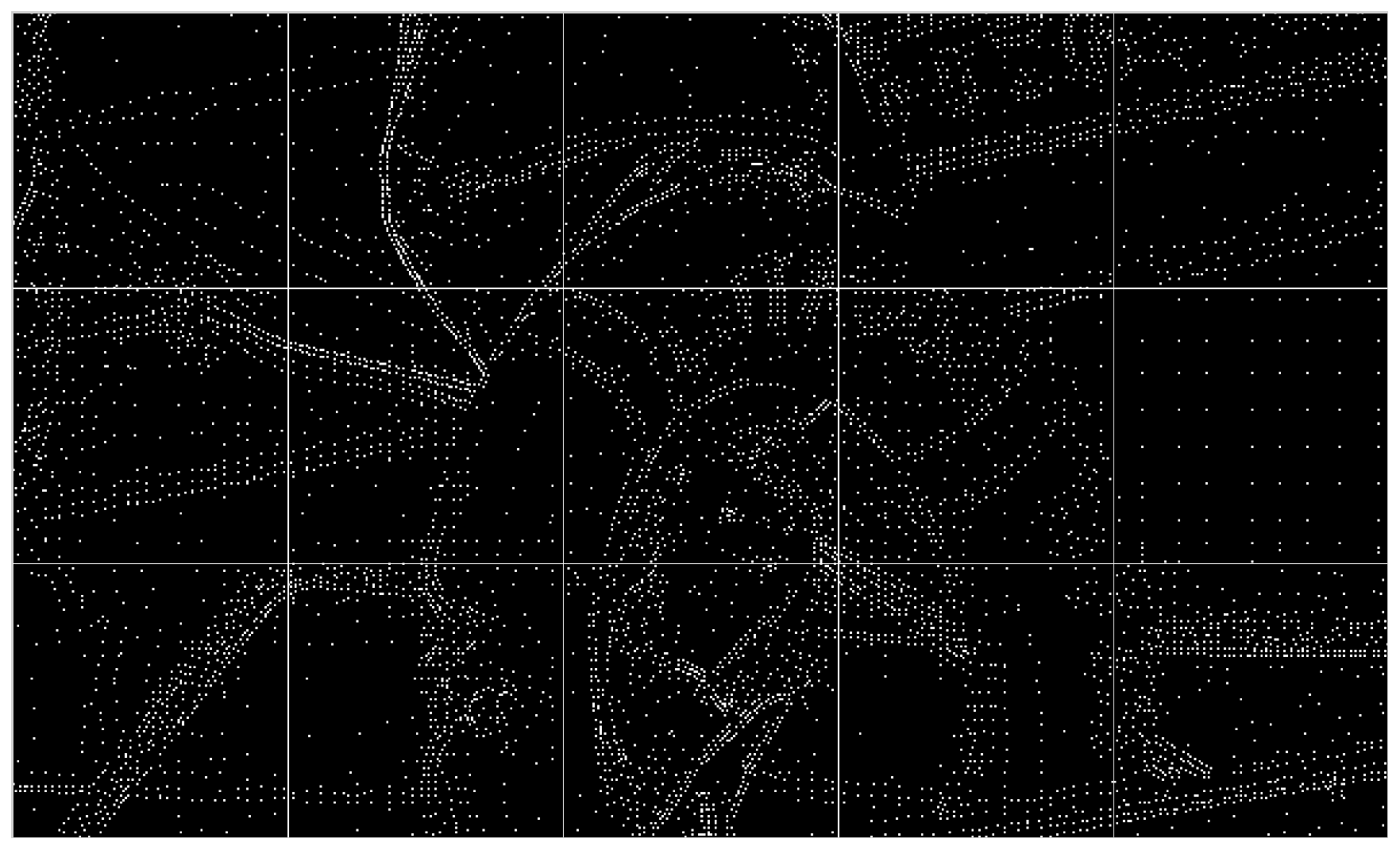} &
\includegraphics[width=6cm]{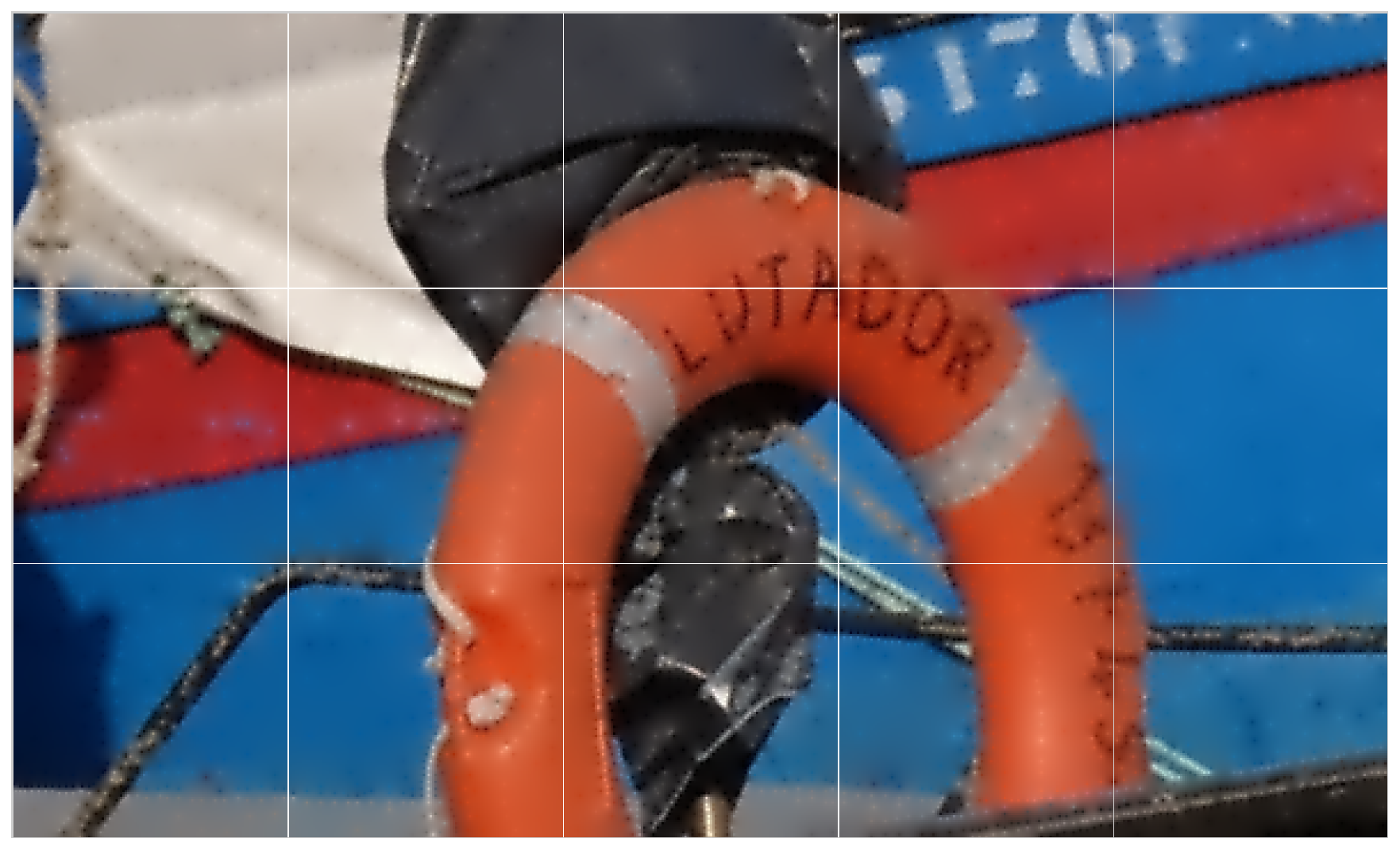} \\
(c) mask & (d) inpainting
\end{tabular}
\caption{\textbf{Stages of the Mask Generation Process.}
(a) Initial image with subdivision into patches. 
(b) Quantised patch densities, and Laplacian magnitude with 
logarithmic dynamic compression for better visibility in the background.
(c) Binary mask with $3\%$ known data.
(d) Inpainted result.
}
\label{fig:patch_based}
\end{figure}

Simply partitioning a large image into patches and generating masks with equal 
density for each leads to suboptimal results: Textured regions
receive too few, while homogeneous areas receive too many 
mask pixels. As such, we require a fast and simple mechanism that estimates 
suitable densities for each patch while taking the whole image into account.

\smallskip
To this end, we estimate a good patch density using the average Laplacian 
magnitude per patch, similar to the approach by Belhachmi et al.~\cite{BBBW08}. 
They have shown that for optimal continuous masks, the local mask density 
should increase with the Laplacian magnitude. The discrete approximation of 
this result through dithering requires significant compromises. 
We, however, aggregate the Laplacian magnitude over an image patch 
to estimate the optimal density and thus avoid dithering completely. 
To the best of our knowledge, we are the first to apply the optimality result 
this way.
This motivates the following algorithm, which is also visualised in Figure 
\ref{fig:patch_based}:
\begin{enumerate}
\item Compute the Laplacian magnitude of the luma channel for every pixel.
\item Rescale it such that its global mean matches the target density.
\item Compute the target patch densities as the mean of the rescaled 
Laplacian magnitude per patch.
Section \ref{sec:inpainting_approx} confirms that these target densities 
correlate well with high-quality masks.
\item Quantise the patch densities to values for which pre-trained mask 
networks are available and generate masks for every patch.
\item Assemble the mask patches into a mask for the whole image.
\end{enumerate}
As an additional benefit, this methodology allows to generate masks with 
arbitrary densities. This is achieved by selecting densities per patch out of 
the available ones such that their mean approximates the desired average well.

\section{Experiments} \label{sec:experiments}
\begin{figure}[t]
\centering
\setlength{\tabcolsep}{1mm}
\begin{tabular}{ccc}
\includegraphics[width=.3\linewidth]{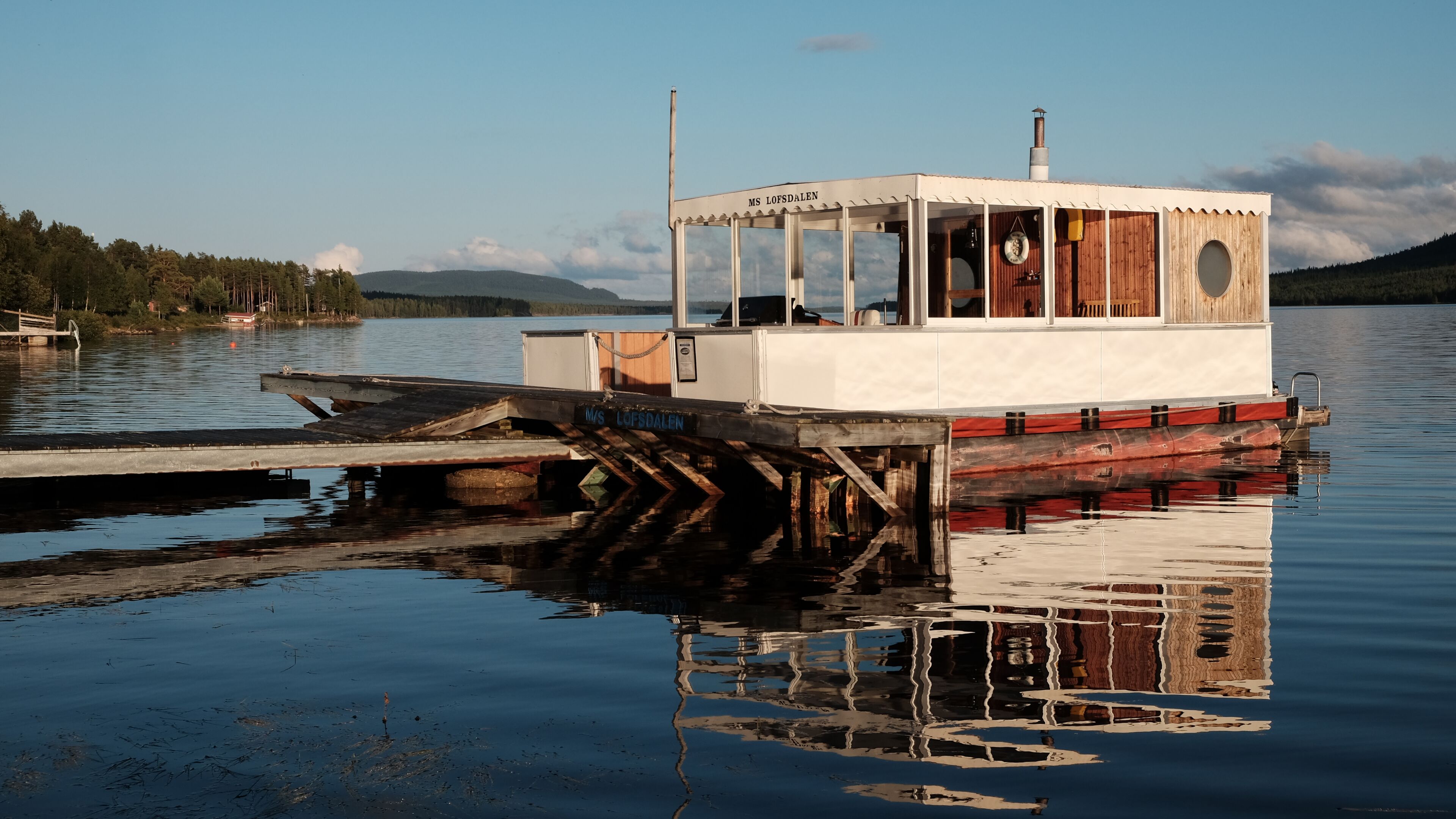} &
\includegraphics[width=.3\linewidth]{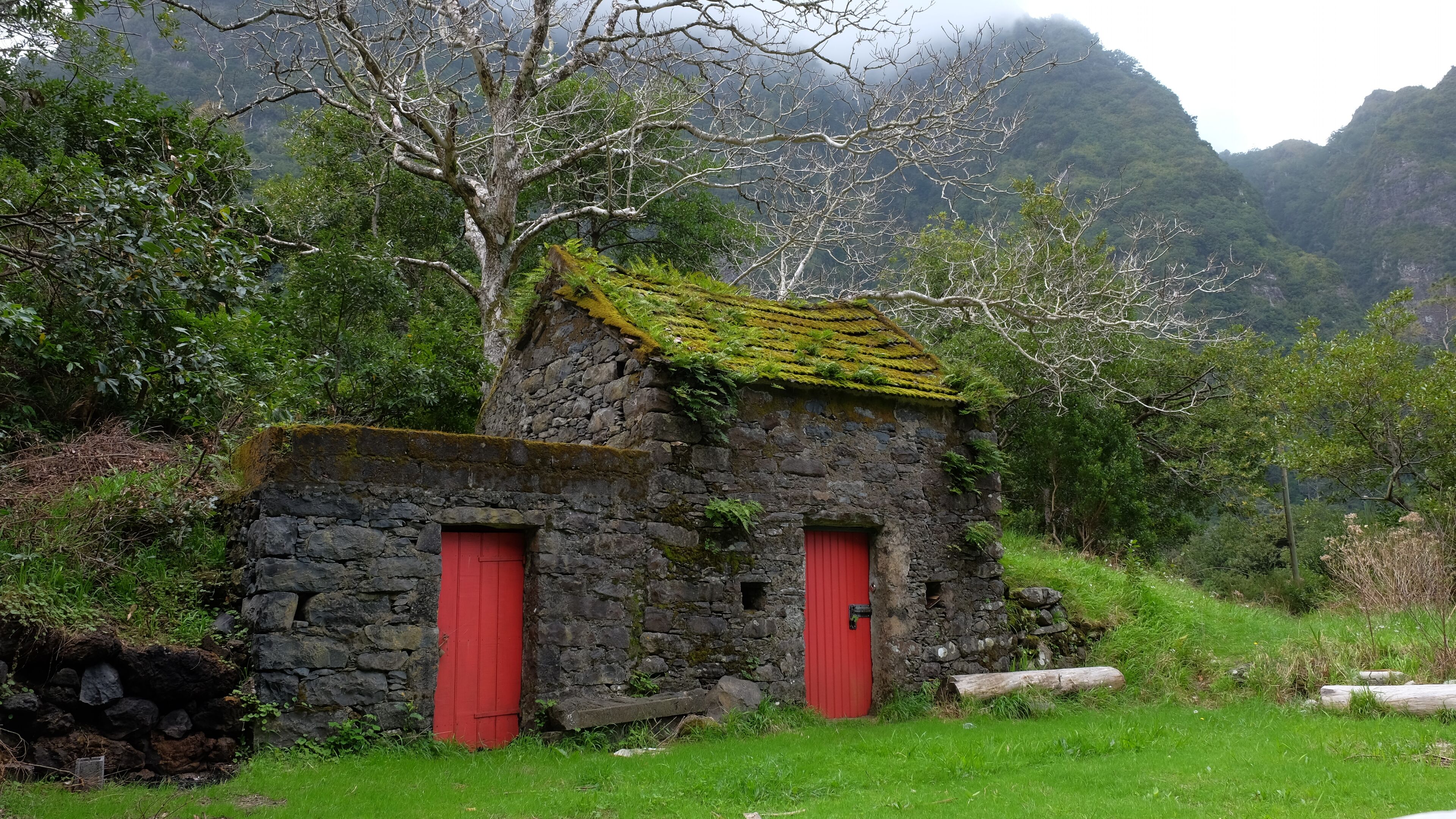} &
\includegraphics[width=.3\linewidth]{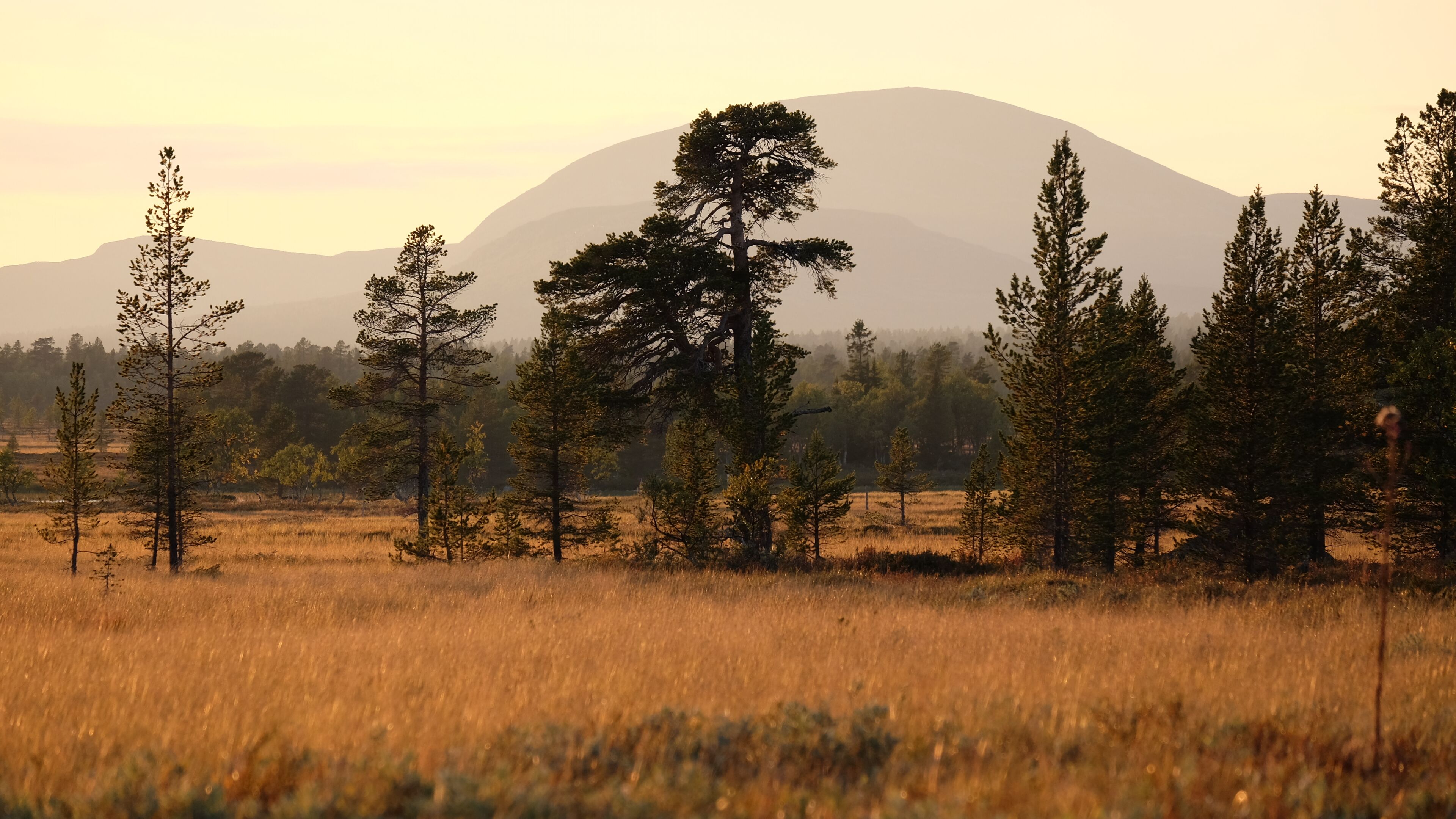} \\[1mm]
\includegraphics[width=.3\linewidth]{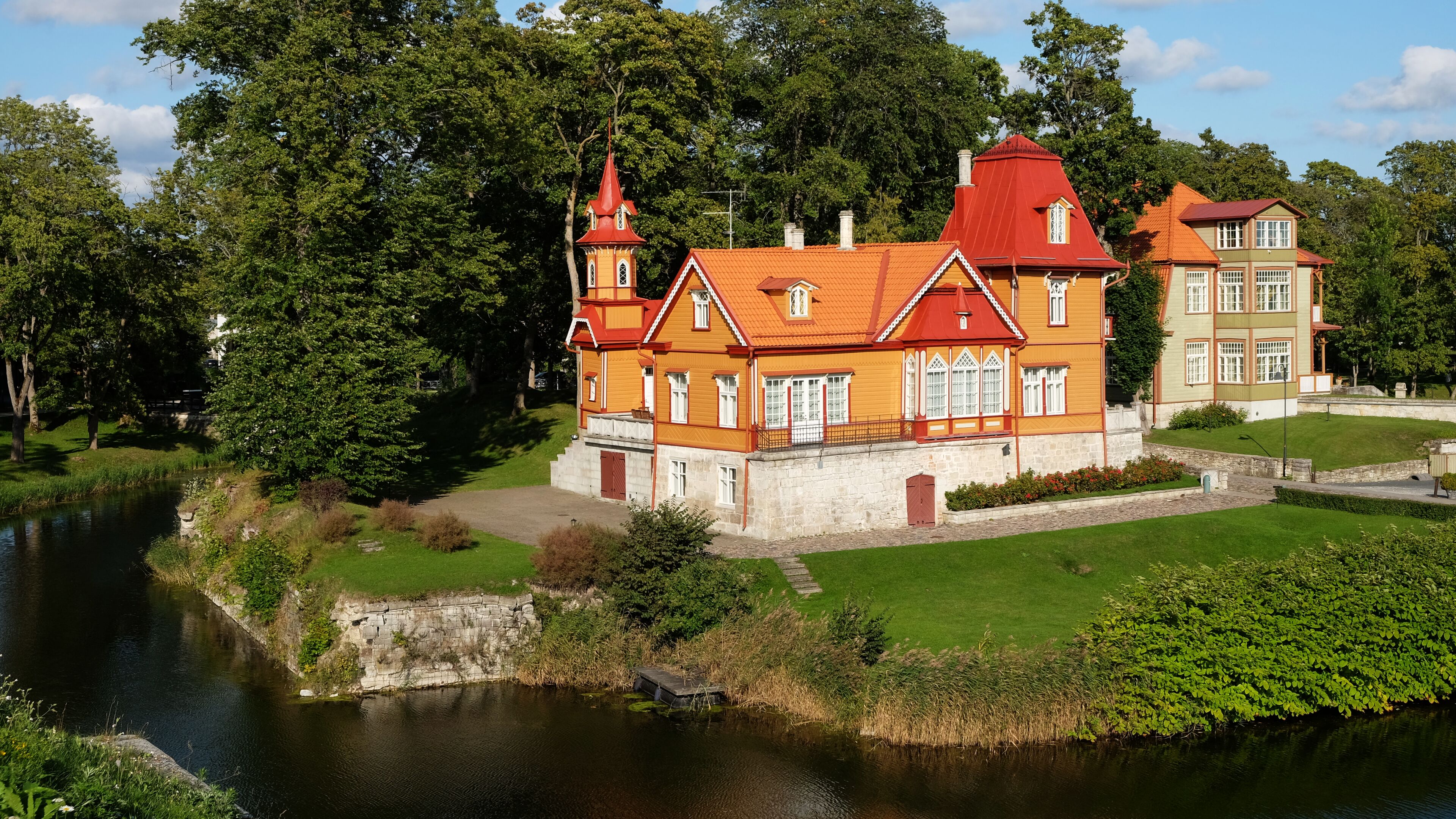} &
\includegraphics[width=.3\linewidth]{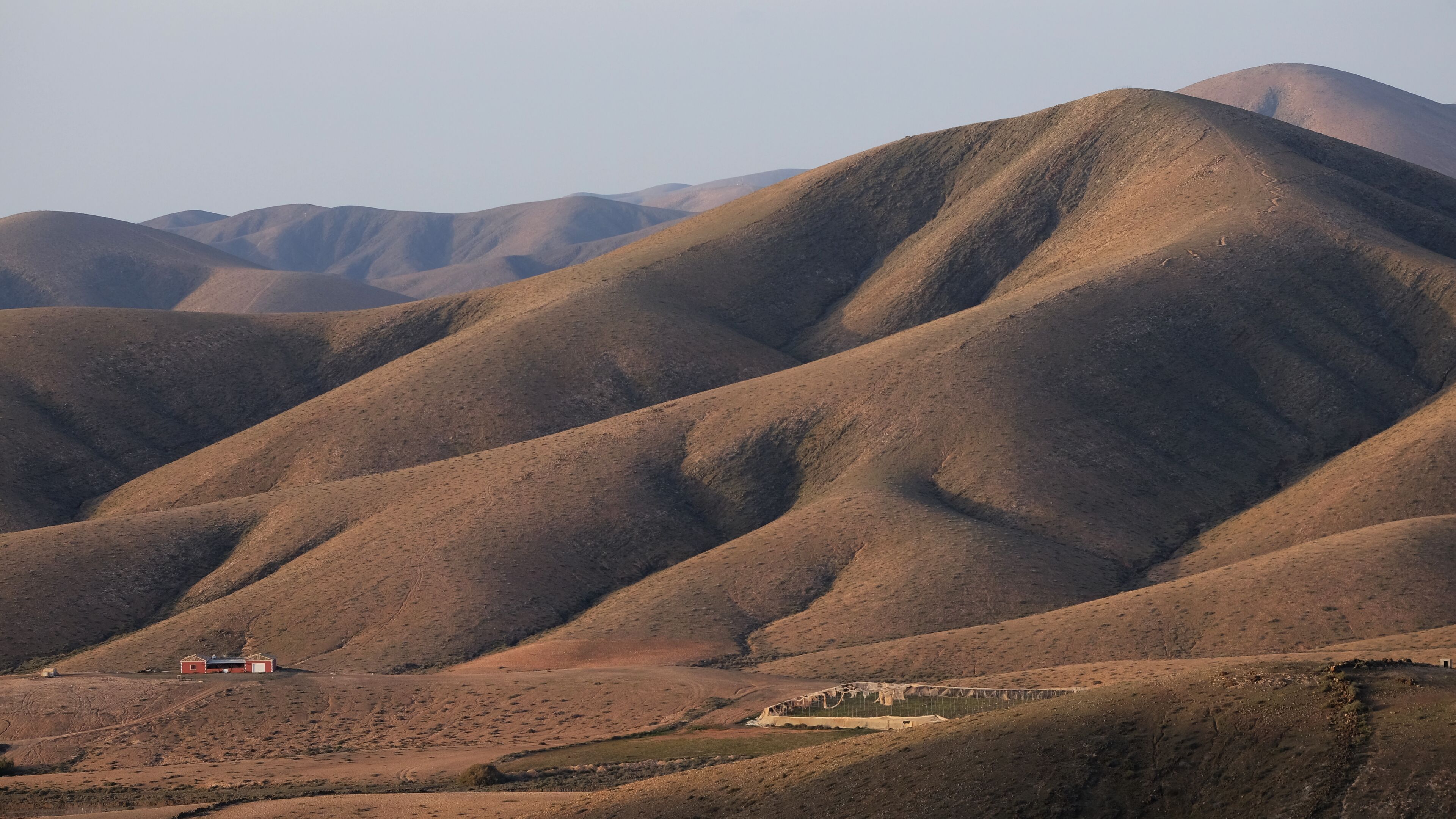} &
\includegraphics[width=.3\linewidth]{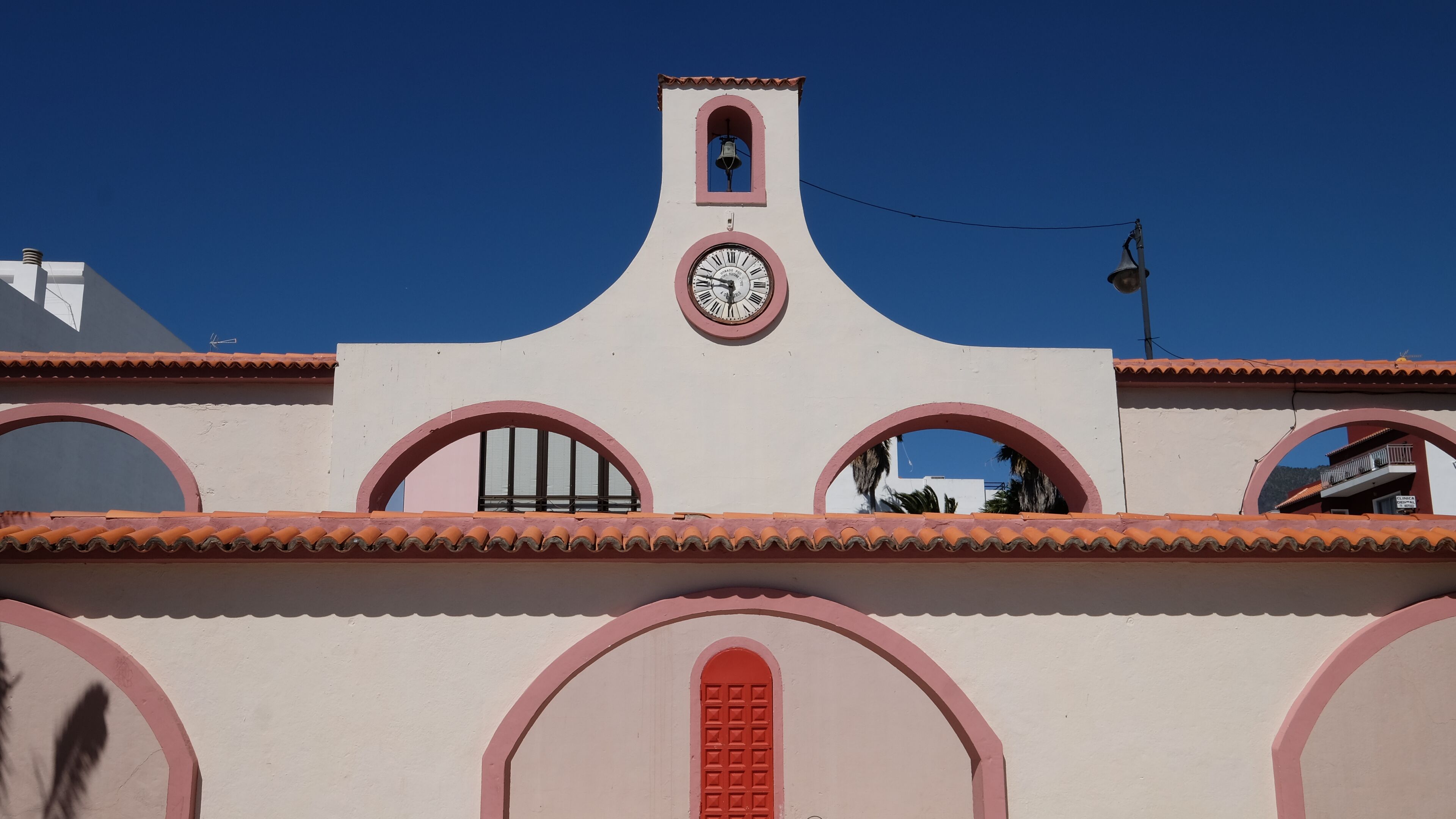} \\[1mm]
\includegraphics[width=.3\linewidth]{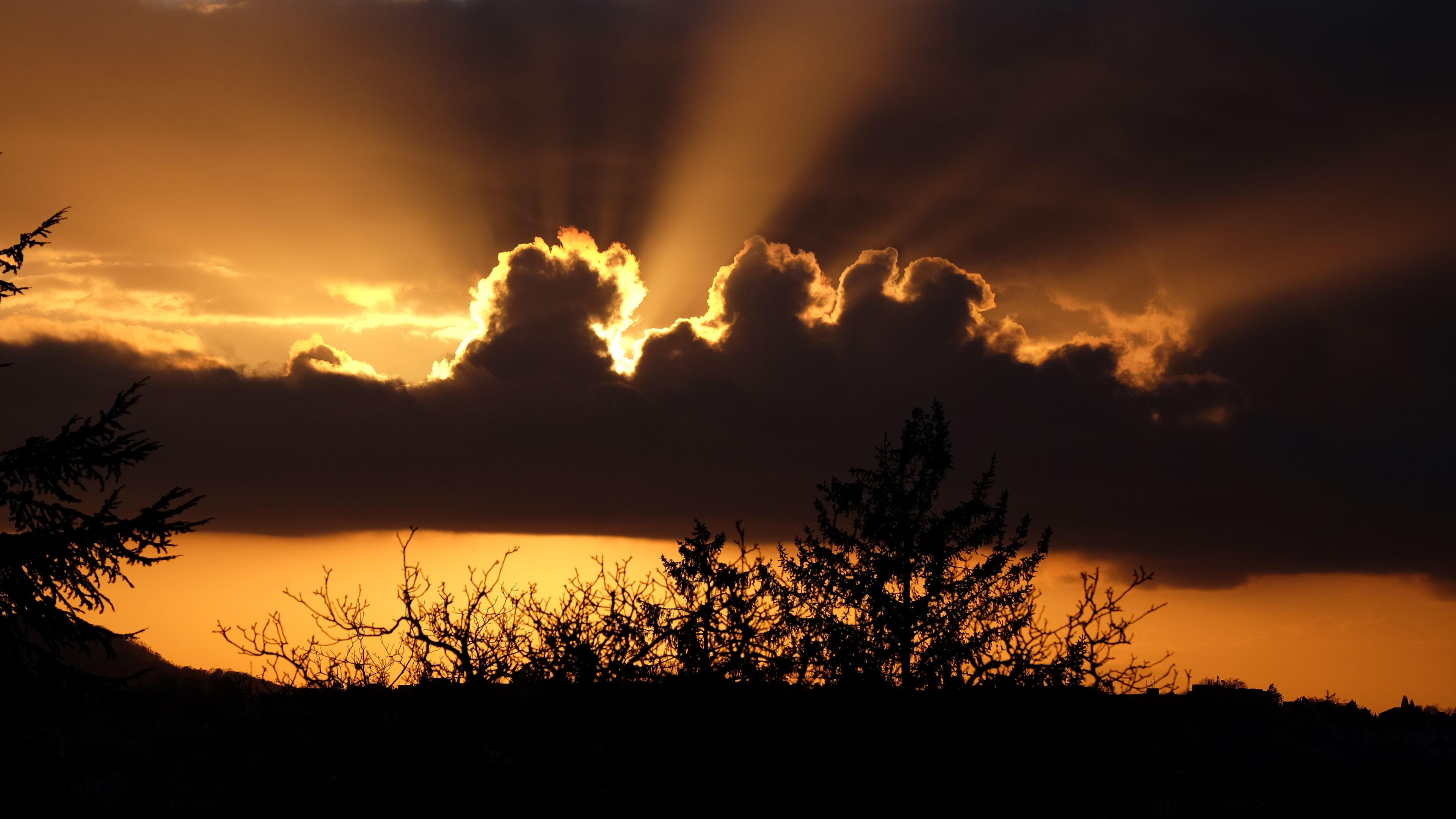} &
\includegraphics[width=.3\linewidth]{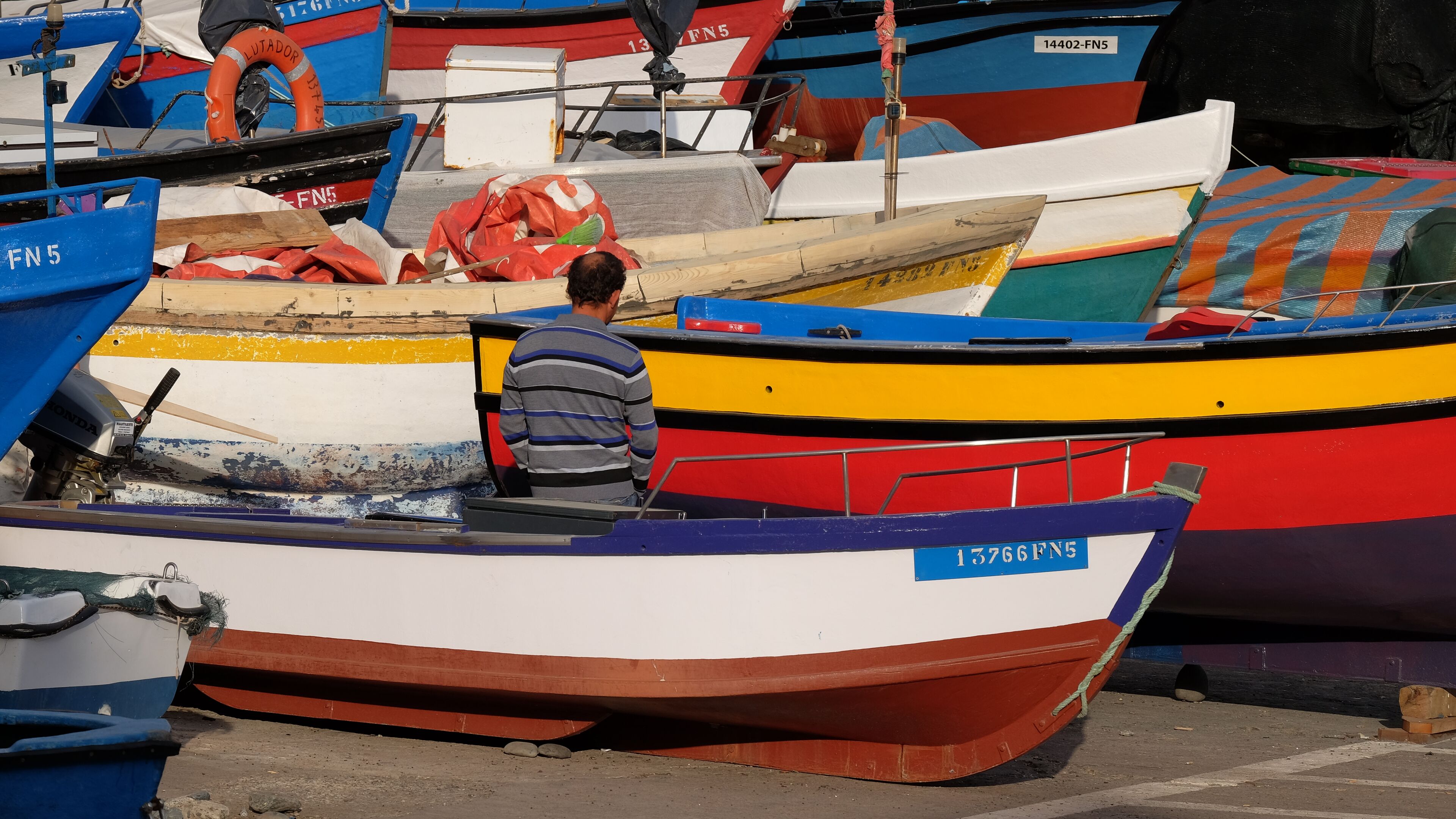} &
\includegraphics[width=.3\linewidth]{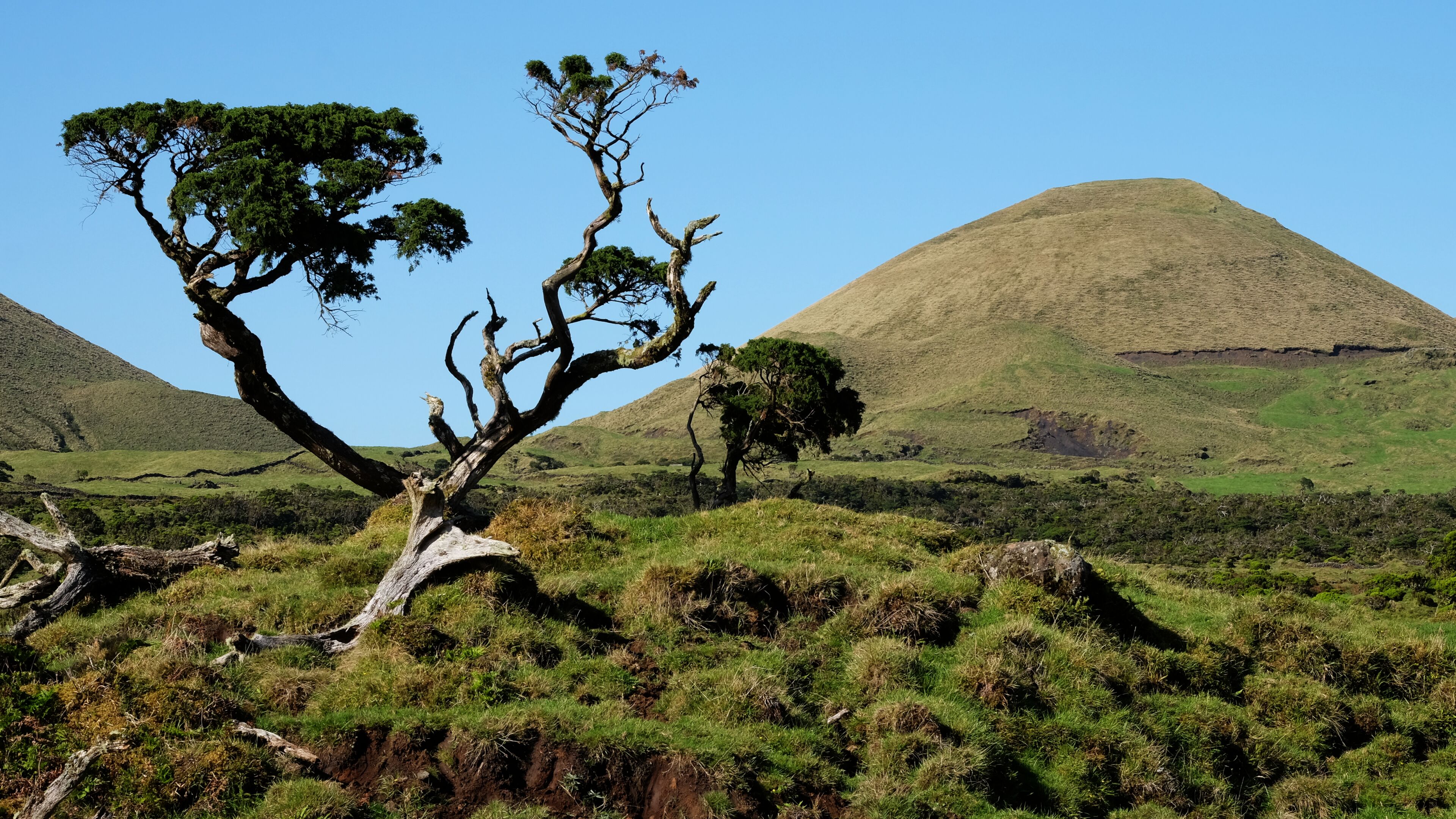} \\[1mm]
\includegraphics[width=.3\linewidth]{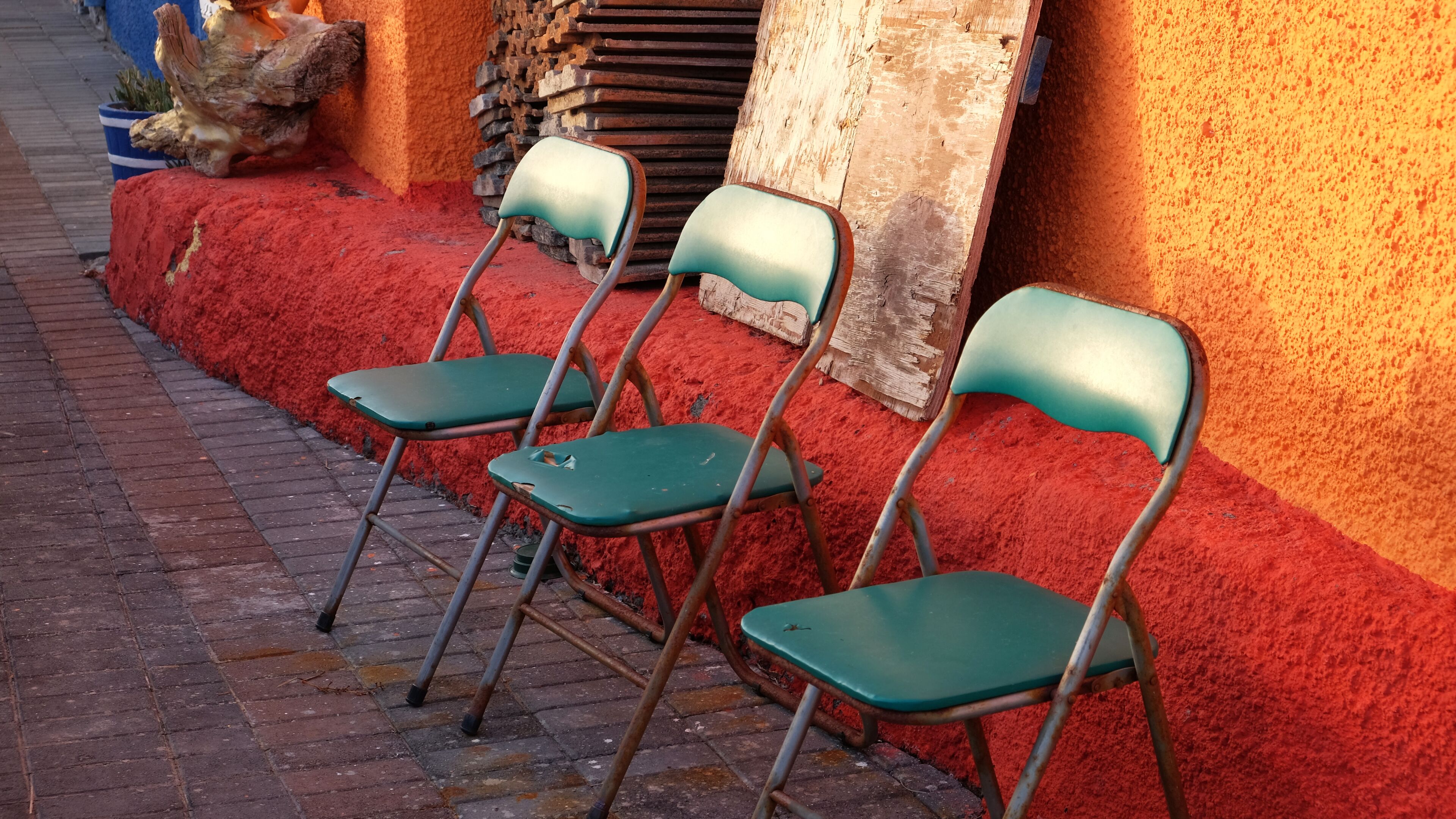} &
\includegraphics[width=.3\linewidth]{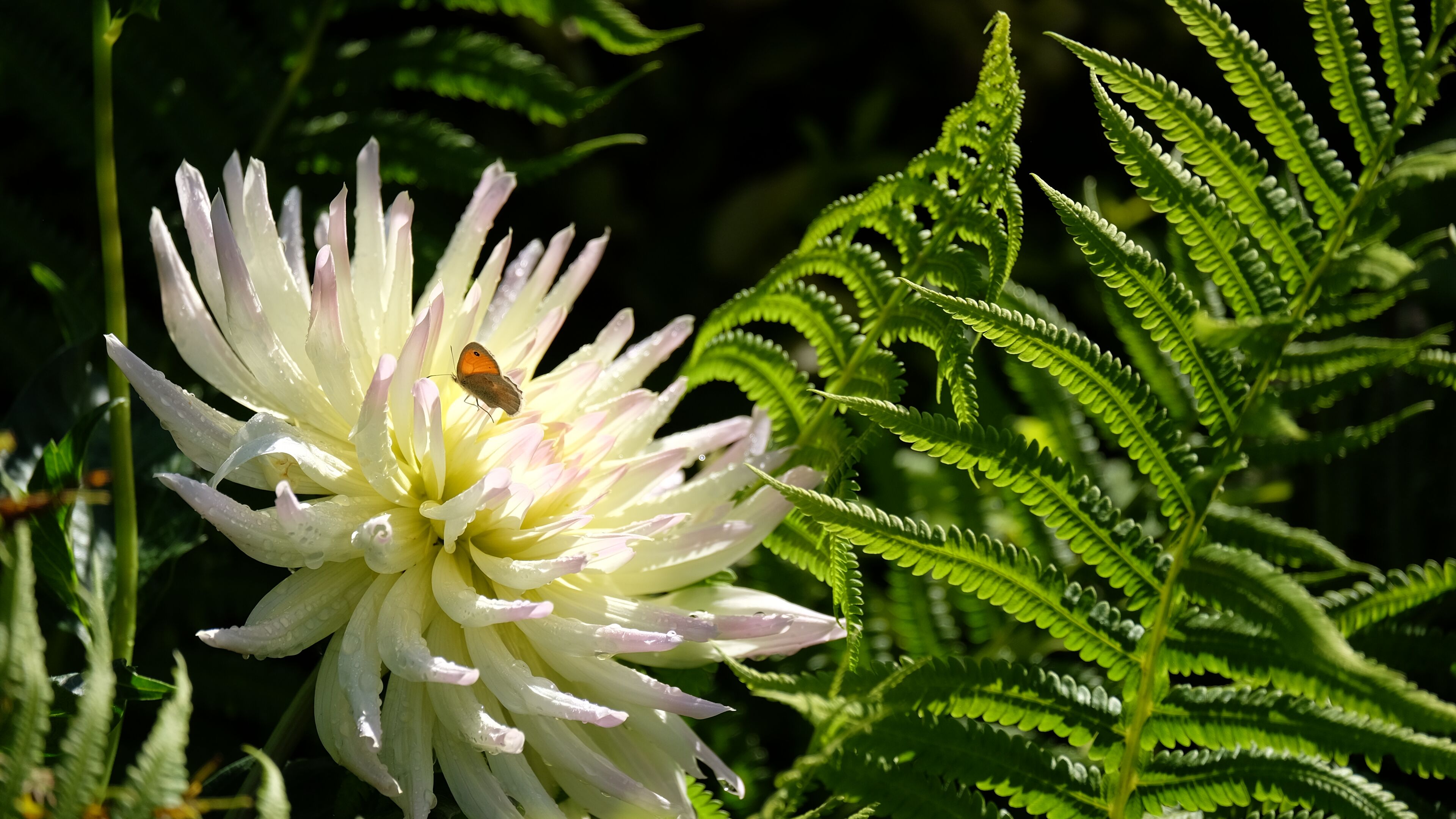} &
\includegraphics[width=.3\linewidth]{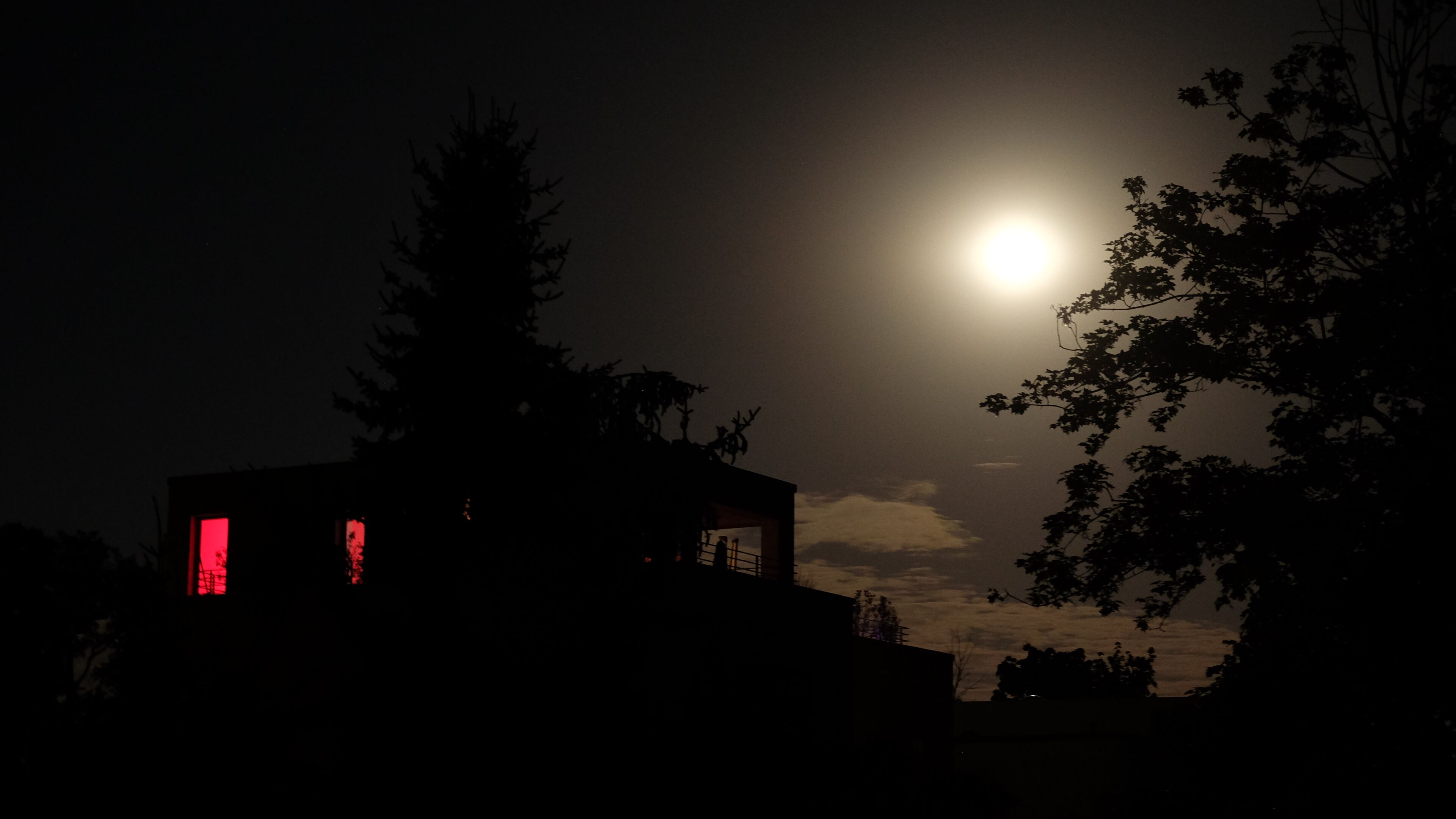} 
\end{tabular}
\caption{Our 12 test images of size $3840 \times 2160$. Photos by J. Weickert.
}
\label{fig:dataset}
\end{figure}
After mentioning the technical details in Section \ref{sec:exp_setup}, 
we show the improvements made by introducing a CG solver into the network 
in Section \ref{sec:inpainting_approx}. 
Additionally, we provide empirical evidence for our patch density estimation in 
Section \ref{sec:dens_approx}, and compare the mask generation performance for 
4K images in Section \ref{sec:4k_exp}.

\subsection{Experimental Setup} \label{sec:exp_setup}
For our mask generator, we use the same small multiscale context 
aggregation network~\cite{VHF21} with 
$\approx2.9$ million parameters and settings as in~\cite{PSAW22} to 
facilitate direct comparisons. 
It is based on a U-Net architecture with four scales and uses blocks of 
parallel dilated convolutions with different dilation rates.
All mask networks are trained with the losses described in Section 
\ref{sec:neural_mask_generation}. The mask variance loss weight is set 
to $\alpha=0.01$.
The surrogate network in Section \ref{sec:inpainting_approx} shares the 
mask network architecture and uses the squared residual of the discrete 
inpainting equation \eqref{eq:discrete_inp} as its loss.
All networks are trained for 100 epochs with a batch size of 8 and the 
Adam optimiser~\cite{KB15} with a learning rate of $5 \cdot 10^{-5}$. 
Performance is evaluated on an AMD Ryzen 7 5800X CPU and an 
Nvidia RTX 3090 GPU. 
For comparisons against~\cite{PSAW22}, we trained on a subset of $100,000$ 
images from ImageNet~\cite{DDSL+09} sampled by Dai et al.~\cite{DCPC19}. Tests 
are performed on the full BSDS500 dataset~\cite{AMFM11}. 
We used $128\times128$ centre crops for both.
The networks used for 4K inpainting were trained on $100,000$ random patches of 
size $120\times120$ from the high-resolution image dataset Div2K~\cite{AT17}. 
Tests are performed on $12$ representative 4K images photographed by one 
of the authors; see Figure~\ref{fig:dataset}. Our selection of pre-trained 
networks is optimised for target densities~$\leq 10\%$, as those are 
practically relevant for compression. The set contains networks for 
different densities in the range from $0.5\%$ to $80\%$. For $1\%$-$15\%$, we 
use increments of $1\%$. For $15\%$-$25\%$ we increment by $2\%$, and by $5\%$ 
for densities between $25\%$ and $50\%$. Finally, in the range of $50\%$-$80\%$ 
we have steps of 10\%.
Reference inpaintings are computed using a CG solver which is stopped after a 
relative decrease of the Euclidean norm of the residual of $10^{-6}$.

\smallskip
For comparisons against model-driven approaches, we use the 
analytic approach~(AA) by Belhachmi et al.~\cite{BBBW08}, and probabilistic 
sparsification~(PS) alone or combined with nonlocal pixel exchange~(PS+NLPE). 
PS uses candidate fractions $p=0.3$ and $q=0.005$. NLPE runs for 5 cycles. The 
other NLPE parameters were optimised individually for the different target 
resolutions. 

\subsection{Quality of Inpainting Approximations} \label{sec:inpainting_approx}
\begin{figure}[t]
\begin{tabular}{cc}
\includegraphics[width=6cm]{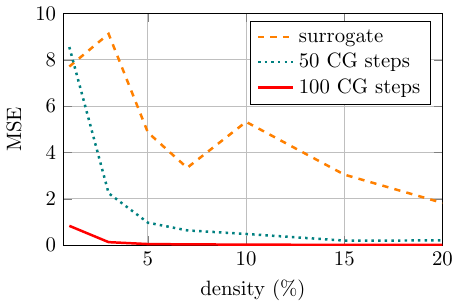} &
\includegraphics[width=6cm]{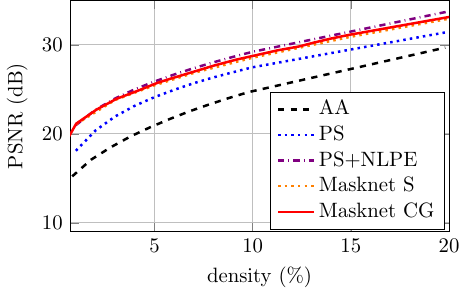} \\
(a) Inpainting Approximation Quality & (b) Mask Network Quality
\end{tabular}
\caption{
\textbf{Comparison of Different Inpainting Approximators.} (a) Distance between 
the inpainting approximations and a converged inpainting.
The MSEs decrease in a nonmonotone fashion due to the varying quality of 
different mask and inpainting networks.
(b) Comparison of mask networks trained with either a surrogate 
inpainter~(Masknet S) or 100 CG iterations (Masknet CG). Even though 100 CG 
iterations are a much better inpainting approximator, the quality of the mask 
network trained with them is only slightly improved. Both methods are 
better than the analytic approach (AA) and probablistic sparsification (PS), 
and are very close to PS with added nonlocal pixel exchange (PS+NLPE).
}
\label{fig:inpainting_approx}
\end{figure}
We measure the quality of different inpainting approximators by comparing 
against inpaintings with our reference solver. To avoid a bias against the 
surrogate inpainting networks, we test on binarised masks generated by their 
corresponding mask networks. 
In Figure \ref{fig:inpainting_approx}(a), we see that 100 CG iterations are a 
better approximator than the surrogate network across all densities. 
Additionally, they allow for faster training: On image batches, one forward and 
backward pass takes only $1.9$ ms per image, while the surrogate requires 
$5.6$ ms.
Nevertheless, the quality of the trained mask network is only slightly improved 
by using CG as an inpainting approximator as can be seen in Figure 
\ref{fig:inpainting_approx}(b). This makes sense as the reconstruction error is 
about $3000$ times larger than the approximation error for $100$ CG steps and 
all densities, making small deviations insignificant.

\begin{figure}[t]
\centering
\includegraphics[width=6cm]{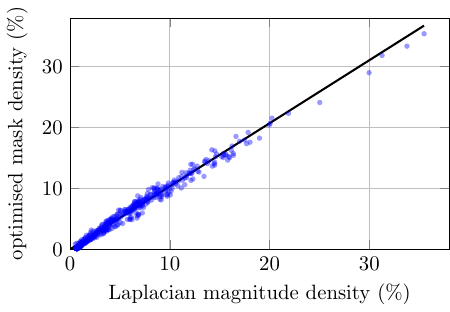} 
\caption{
\textbf{Patch Density Correlation between Laplacian Magnitude 
and High Quality Mask.} 
Comparison of the mean rescaled Laplacian magnitude for each patch of the image 
\textit{lofsdalen} against patch densities of a high quality mask, both with a 
total density of $5\%$. A correlation coefficient of $0.994$ confirms the 
strong connection.
}
\label{fig:densitycorrelation}
\end{figure}

\subsection{Justification of Mask Pixel Distribution} \label{sec:dens_approx}
In Figure~\ref{fig:densitycorrelation} we show that the rescaled Laplacian 
magnitude is a good predictor of optimal mask densities at a patch level.
There we compare against the patch densities of a well optimised mask 
generated with the approach from Chizhov and Weickert~\cite{CW21}. 
The relationship between the patch densities is almost linear, and the mean 
absolute error between densities is only $0.5\%$. 
While dithering the rescaled Laplacian magnitude leads to low-quality masks, 
its aggregation over a patch avoids this and produces a good coarse scale 
density estimate.

\subsection{High-Resolution Mask Generation} \label{sec:4k_exp}
\begin{figure}[t]
\begin{tabular}{cc}
\includegraphics[width=5.9cm]{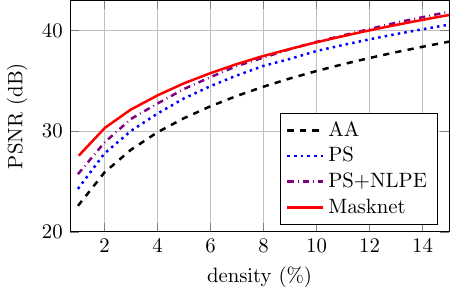} &
\includegraphics[width=6.1cm]{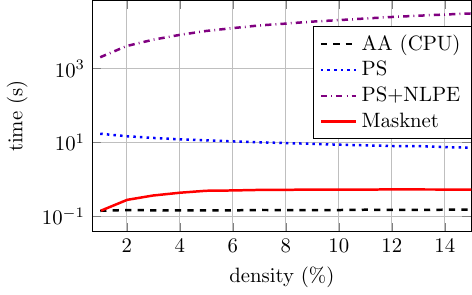} \\
(a) quality  & (b) runtime
\end{tabular}
\caption{
\textbf{Spatial Optimisation.} Our masknet outperforms the faster analytic 
approach,
but also the slower PS across all 
densities. It even beats the significantly slower PS+NLPE for 
densities $\leq 8\%$. Time measurements exclude input/output operations.
}
\label{fig:w4k_performance}
\end{figure}
\begin{figure}[t]
\centering
\begin{tabular}{cc}
image \textit{lofsdalen} & zoom to region \\
\includegraphics[width=0.41\linewidth]{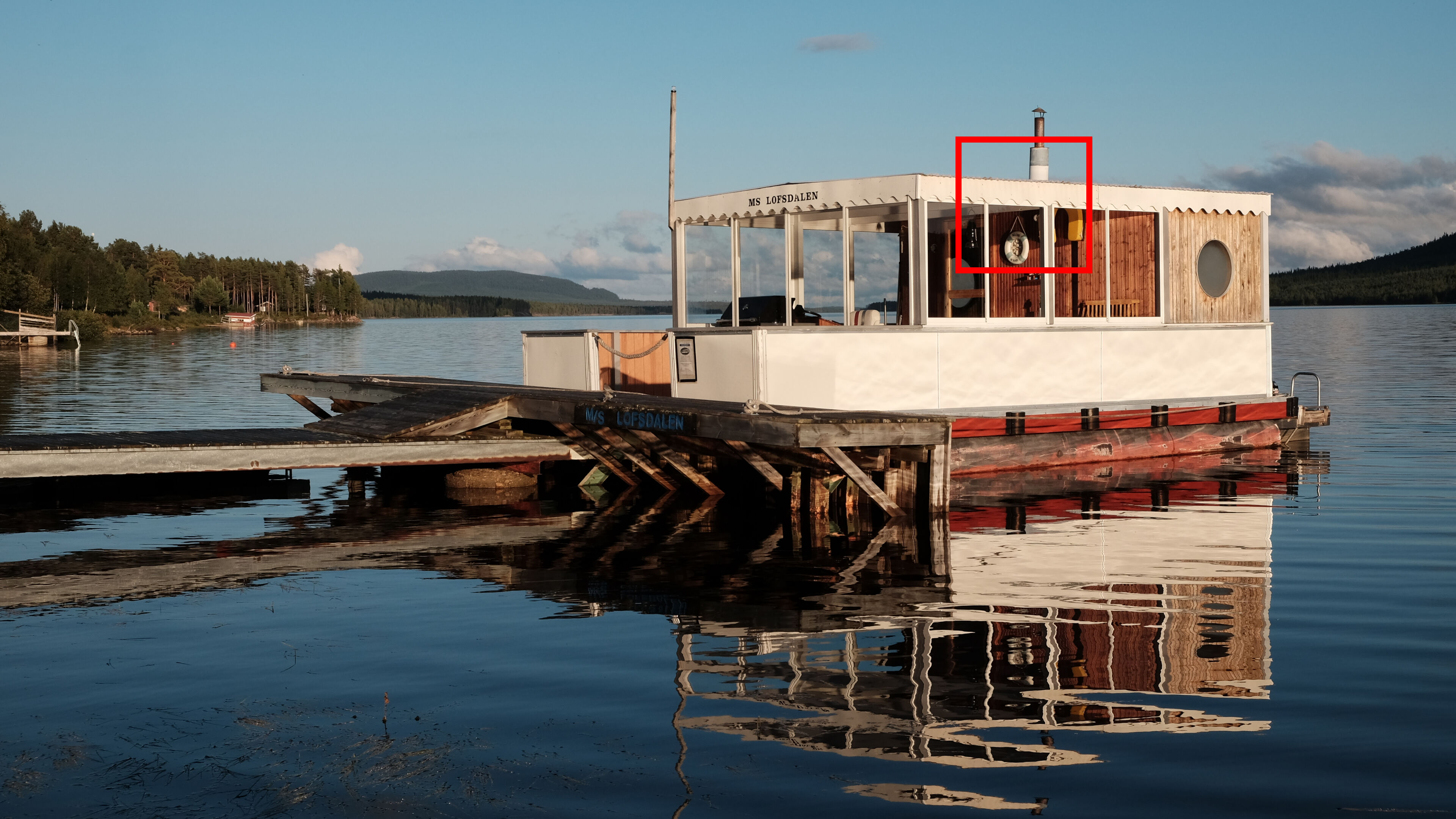}
\hspace{1mm} &
\includegraphics[width=0.23\linewidth]{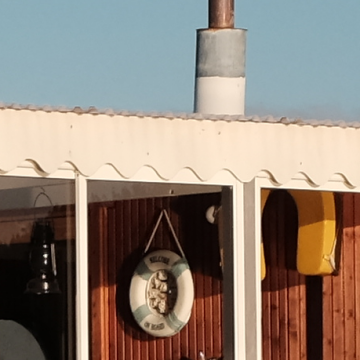} 
\end{tabular}
\\[1mm]
\setlength{\tabcolsep}{1mm}
\begin{tabular}{ccccc}
& AA~\cite{BBBW08} & PS~\cite{MHWT12} & PS+NLPE~\cite{MHWT12} & our masknet \\
\rotatebox{90}{\hspace{0.8cm} mask} & 
\includegraphics[width=0.22\linewidth]
  {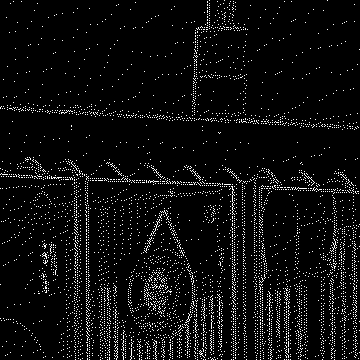} &
\includegraphics[width=0.22\linewidth]
  {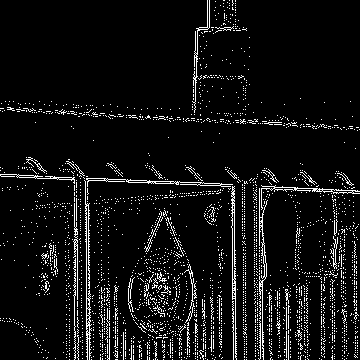} &
\includegraphics[width=0.22\linewidth]
  {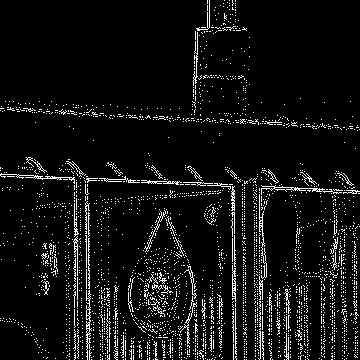} &
\includegraphics[width=0.22\linewidth]
  {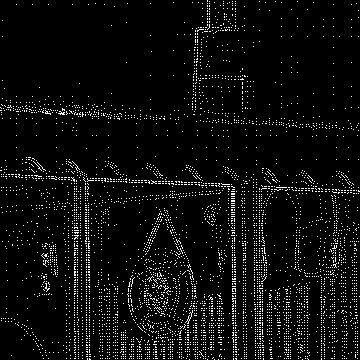} \\[1mm]
\rotatebox{90}{\hspace{0.55cm} inpainted} & 
\includegraphics[width=0.22\linewidth]
  {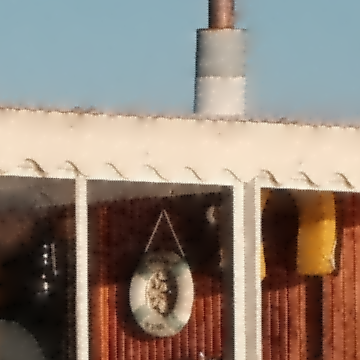} &
\includegraphics[width=0.22\linewidth]
  {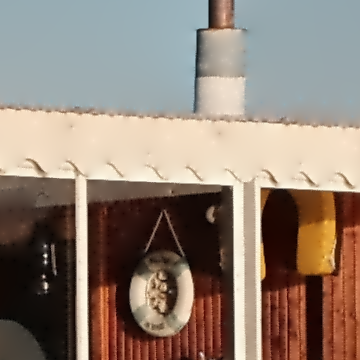} &
\includegraphics[width=0.22\linewidth]
  {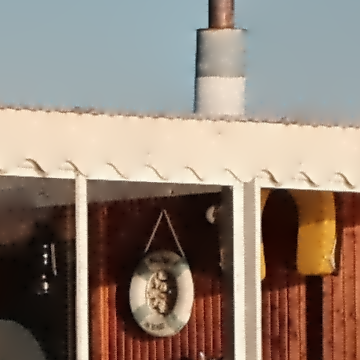} &
\includegraphics[width=0.22\linewidth]
  {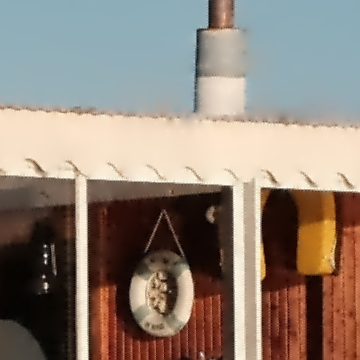} \\[-0.5mm]
& PSNR: 28.86 & PSNR: 29.32 & PSNR: 30.03 & PSNR: 32.18 
\end{tabular}\vspace{-1mm}
\caption{\textbf{Visual Comparison for 4\% Mask Density on \textit{lofsdalen}.} 
PSNRs are for the whole image. Notice the discoloured sky for PS and PS+NLPE. 
The full images are available in the supplementary material. 
\label{fig:visual_comp}}
\end{figure}

The tests on 4K images in Figure~\ref{fig:w4k_performance}(a) show that our 
masknet trained with 100 CG iteration is superior to AA and PS for all 
densities. In addition, we are even able to outperform PS+NLPE on densities 
smaller than $8\%$. This range is practically relevant for inpainting-based 
 compression.

\smallskip
In Figure~\ref{fig:w4k_performance}(b) we compare the speed of the different 
methods. 
The runtime of PS+NLPE scales with the required inpaintings, taking 
between $30$ minutes and $8.5$ hours depending on the density. 
In contrast, our neural approach takes only about $0.6$ seconds per image.
It is even quicker for the lowest densities where fewer of the pre-trained mask 
networks are active and the batches per net are larger.
Our masknet is also more than $10$ times faster than the qualitatively 
worse PS. The analytic approach requires no inpaintings, and even 
a CPU-based implementation is significantly faster than our neural method.
Still, its quality is inferior by a large margin.
In Figure \ref{fig:visual_comp}, the most noticeable differences are slightly 
blurry edges for AA, and a discoloured sky for PS and PS+NLPE. 

\section{Conclusions} \label{sec:conclusion}
Our coarse-to-fine approach for mask generation is orders of magnitude 
faster than qualitatively similar approaches on 4K images. This shows that the 
estimation of patch densities using the optimality result by 
Belhachmi et al.~\cite{BBBW08} works well, and our mask networks are able to 
outperform dithering. Performance of the neural approach relative to PS+NLPE 
even grew with the increase in resolution. 
Our experiments suggest that the coarse-to-fine approach produces local 
problems that are easier and faster to solve in high quality. 
This is in line with transform-based codecs like JPEG~\cite{PM92},
which also use a block structure to enable parallelisation and reduce 
complexity.  

\smallskip
In the future, we will investigate these coarse-to-fine approaches for mask 
generation further using various local solvers.
This avenue of research becomes even more desirable given the ever-increasing 
parallelisation capabilities of modern GPUs, rapidly increasing 
image resolutions, and the fact that reasonably simple solvers such as 
CG cannot offer optimal linear complexity.

\bibliographystyle{splncs04}
\bibliography{myrefs.bib}

\end{document}